\documentclass{article}
\usepackage{amssymb}

\newtheorem{theorem}{Theorem}
\newtheorem{acknowledgement}[theorem]{Acknowledgement}

\input{tcilatex}
\begin{document}

\title{On quantum Hall effect: Covariant derivatives, Wilson lines, gauge
potentials, lattice Weyl transforms, and Chern numbers}
\author{Felix A. Buot \\
%EndAName
CTCMP, \ Cebu Normal University, Cebu City 6000\\
Philippines,\\
C\&LB Research Institute, Carmen, Cebu 6005\\
Philippines,\\
LCFMNN, University of San Carlos, Cebu City 6000,\\
Philippines}
\maketitle

\begin{abstract}
We show that the gauge symmetry of the nonequilibrium quantum transport of
Chern insulator in a uniform electric field is governed by the Wilson line
of parallel transport operator coupled with the dynamical translation
operator. This is dictated by the minimal coupling of derivatives with gauge
fields in $U\left( 1\right) $ gauge theory. This parallel transport symmetry
consideration leads to the integer quantum Hall effect in electrical
conductivity obtained to first-order gradient expansion of the
nonequilibrium quantum transport equations.
\end{abstract}

\section{Introduction}

In previous papers \cite{previous, comments, jagna}, we make use of the
gapped energy-band structure of solids under external electric field to
derive the integer quantum Hall effect (IQHE) of Chern insulator. We employ
the real-time superfield and lattice Weyl transform nonequilibrium Green's
function (SFLWT-NEGF) \cite{buot8} quantum transport formalism \cite{buot1,
bj} to the first-order gradient expansion to derive the topological Chern
number of the IQHE for two-dimensional systems, as an integral multiple of
quantum conductance, also known as the minimal contact conductance in
mesoscopic physics \cite{buot8}.

We find that the quantization of Hall effect occurs strictly not to first
order in the electric field \textit{per se} but rather to first-order
gradient expansion in the nonequilibrium quantum transport equation. The
Berry connection and Berry curvature is the fundamental physics \cite{fego}
behind the exact quantization of Hall conductance in units of $\frac{e^{2}}{h%
}$, which also happens to coincide with the source and drain \textit{contact}
conductance per spin in a closed circuit of mesoscopic quantum transport 
\cite{buot8}.

In Ref.\cite{previous}, we have shown that the $\left( p.q;E.t\right) $
phase-space is renormalized to that of $\left( \mathcal{\vec{K}},\mathcal{E}%
\right) $ phase-space, where%
\begin{equation}
\mathcal{\vec{K}}=\vec{p}+e\vec{F}t,  \label{eq1-1}
\end{equation}%
and 
\begin{equation}
\mathcal{E}=E_{0}+e\vec{F}\cdot \vec{q}  \label{eq2-2}
\end{equation}%
Here, the uniform electric field, $\vec{F}$, is in the $x$-direction, and
the Hall current is in the $y$-direction.

We have identified the topological invariant in $\left( \mathcal{\vec{K}},%
\mathcal{E}\right) $ phase-space nonequilibrium quantum transport equation
leading to IQHE. Moreover, the formula is also applicable to gapped
Landau-level structure of a free electron gas in intense magnetic field \cite%
{kdp} since the variable $\mathcal{\vec{K}}$ can incorporates the external
vector potential, and its corresponding parallel transport, if present. Note
that the change of variables from $\mathcal{\vec{K}}$ to $\vec{p}$ in the
integration over the whole Brillouin zone has a Jacobian unity. It was shown
in previous paper \cite{previous} that the correct expression for the IQHE
conductivity given by%
\[
\ \sigma _{yx}=\frac{e^{2}}{h}\sum\limits_{\alpha }\frac{\Delta \phi _{total}%
}{2\pi }=\sum\limits_{\alpha }\frac{e^{2}}{h}n_{\alpha } 
\]

In the present paper, we show that Eqs, (\ref{eq1-1}) and (\ref{eq2-2})
directly arise from the \textit{local} gauge symmetry under a uniform
electric field of localized Wannier functions centered on lattice sites.
These follow from covariant derivatives leading to the dynamical
spatio-temporal finite translation operators that is coupled with the Wilson
line parallel-transport operators.

\section{Translation operators in uniform electric fields}

In gauge theory, the minimally-coupled gauge covariant derivative is defined
as%
\[
D_{\mu }=\partial _{\mu }+i\alpha ^{\mu }A_{\mu }
\]%
where $A_{\mu }$ is the electromagnetic four "vector" potential, and $\alpha
^{\mu }$ is the serve as a coupling "charge". In electrodynamics, $\alpha
^{x,y,z}=\frac{e}{\hbar c}$, whereas $\alpha ^{0}=\frac{e}{\hbar }$.
Therefore a finite translation operator $T\left( q_{\mu }\right) $ by the
four coordinates $q_{\mu }$ can be written as%
\[
T\left( q_{\mu }\right) =\exp \left[ q_{\mu }\left( D_{\mu }\right) \right]
=\exp \left[ q_{\mu }\left( \partial _{\mu }+i\alpha ^{\mu }A_{\mu }\right) %
\right] 
\]%
where we have used the Einstein summation convention.

\subsection{Wilson lines}

The part of $T\left( q_{\mu }\right) $ given by $\exp \left[ q_{\mu }\left(
i\alpha ^{\mu }A_{\mu }\right) \right] $ represent the Wilson lines. In
general, for non-Abelain gauge theories, the Wilson line is often written
as, 
\[
W_{C}=\mathcal{P}\exp \left[ i\dint\limits_{q=0}^{q_{\mu }}A_{\mu }dx^{\mu }%
\right] 
\]%
where $\alpha $'s are absorbed in $A_{\mu }$, $\mathcal{P}$ is the path
ordering operator since the $A_{\mu }$'s are generally non-Abelian and do
not commute, e.g., like the Pauli matrices.

The Wilson loop is defined as the average of the trace of Wilson lines
around a closed loop, where the average is taken using the Chern-Simons
Lagrangian which we will not go into. Moreover, we will not be dealing with
non-Abelian $A_{\mu }$'s and our Wilson lines are explicitly gauge
invariant, being expressed in terms of the gauge-field strength.

The importance of the Wilson line as a parallel transport operator in the
gauge independent formulation of gauge theories has been emphasized by
Mandelstam \cite{mandel} and further developed by Wu and Yang \cite{wu}. The
term given by $\dint\limits_{L}A_{\mu }dx^{\mu }$ is the electromagnetic
Berry phase. Precisely, the definition of the Berry phase $\phi $ reads 
\begin{eqnarray}
\exp i\phi \ \left\vert \psi \left( x+q_{x},t+T\right) \right\rangle 
&=&T\left( q_{\mu }\right) \left\vert \psi \left( x,t\right) \right\rangle
=\exp \left[ q_{\mu }\left( \partial _{\mu }+i\alpha ^{\mu }A_{\mu }\right) %
\right] \left\vert \psi \left( x,t\right) \right\rangle   \nonumber \\
&=&\exp \left\{ \left[ q_{x}\cdot \partial _{x}+T\partial _{t}\right] +i\phi
\right\} \left\vert \psi \left( x,t\right) \right\rangle   \nonumber \\
&=&\exp \left\{ \frac{i}{\hbar }\left[ \dint\limits_{0}^{q_{x}}\hat{P}\cdot
dx-i\mathcal{H}\left( t^{\prime }\right) dt^{\prime }\right] +i\phi \right\}
\left\vert \psi \left( x,t\right) \right\rangle   \label{peierls}
\end{eqnarray}%
where $\hat{P}$ is the momentum operator, $\mathcal{H}$ is the energy
operator, $A_{\mu }$ is the electromagnetic Berry connection. The last line
serves as a generalization in terms of quantum mechanical operators, $\hat{P}
$ and $\hat{H}$. The extra phase acquired in translation given by 
\begin{equation}
\phi =\dint \alpha ^{\mu }A_{\mu }dx^{\mu }  \label{WL}
\end{equation}%
is the geometrical phase. In this paper, it is identified with the
generalized \textit{Peierls phase factor} in energy band dynamics of solid
state physics. In quantum physics, the adiabatic contour integral of Eq. (%
\ref{WL}) is now generally known as the Berry phase and $A_{\mu }$ is now
generally known mathematically as the \textit{connection}.

\subsubsection{Peierls phase factor}

The factor $\exp i\phi $ on the left side of Eq. (\ref{peierls}) is the
Peierls-phase factor well-known in energy band dynamics of solids \cite%
{wannier}.

\subsection{Gauge potential in uniform electric fields}

In this section, we determined the gauge potential $A_{\mu }=\left(
A_{0},A_{x},A_{y},A_{z}\right) =\left( A_{0},\vec{A}\right) $, where $A_{0}$
is the time component. We have%
\[
\vec{A}=\vec{F}ct 
\]%
where $\vec{F}$ is the uniform electric field. Clearly, $\alpha ^{\mu
}A_{\mu }=\frac{e}{\hbar c}\vec{A}$ has the same units as $\vec{\partial}$
and $\frac{e}{c}\vec{A}$ has the same dimensional units as the momentum,
i.e., 
\[
\hat{P}+\frac{e}{c}\vec{A} 
\]%
represents the minimal coupling in quantum electrodynamics.

For $A_{0}$, we have 
\begin{equation}
\alpha ^{0}A_{0}=-\frac{1}{\hbar }e\vec{F}\cdot \vec{q}  \label{signconv}
\end{equation}%
has the same units as $\partial _{t}$ and hence from Eq. (\ref{peierls}), we
have 
\[
\frac{1}{\hbar }\left( E_{o}+e\vec{F}\cdot \vec{q}\right) 
\]%
represents the coupling of the zero-field energy, $E_{o}$, to the electric
field at arbitrary lattice site $\vec{q}$. Note that the negative sign of
Eq. (\ref{signconv}) is dictated by the equation for the force as the
negative gradient of the potential, namely, 
\[
e\vec{F}=-\vec{\nabla}\left( -e\vec{F}\cdot \vec{q}\right) 
\]

Therefore under uniform electric fields, we have the Wilson lines for our
Abelian gauge potential given by 
\begin{eqnarray*}
W_{C} &=&\exp \left[ i\dint \alpha ^{\mu }A_{\mu }dx^{\mu }\right] \\
&=&\exp \left[ i\left( \alpha ^{0}A_{0}t+\alpha \vec{A}\cdot q\right) \right]
\\
&=&\exp \left[ i\left( \left\{ -\frac{1}{\hbar }e\vec{F}\cdot \vec{q}%
\right\} \ t+\left\{ \frac{e}{\hbar c}\vec{F}ct\right\} \cdot q\right) %
\right]
\end{eqnarray*}%
Thus, the Wilson lines serves as the generalization of Peierls phase factor
in solid state physics.

\section{Alternative derivation of gauge potentials and Peierls phase factor}

In contrast to the derivation given above, which directly use the covariant
derivative in $U\left( 1\right) $ gauge theory, here we will use a
self-consistent Heisenberg equation for the spatio-temporal displacement
operators to determine the gauge potentials and hence the generalized
Peierls phase factor. Remarkably, the self-consistent Heisenberg equation of
motion automatically fixes the negative sign of the scalar gauge of Eq. (\ref%
{signconv}).

Note that \textit{finite} displacement operators are characteristically
exponential operators amenable to Fourier series expansion. Remarkably,
phase factors are generally acquired due to displacement or motion in
parameter space in the presence of electromagnetic fields. This is
ubiquitous in solid-state physics (e.g., Peierls phase factor in magnetic
fields) before the Berry connection become mainstream and fashionable. To
begin, we write our bare Hamiltonian of electrons as 
\[
H=H_{o}-e\vec{F}\cdot \vec{r}, 
\]%
where $H_{o}$ is the periodic Hamiltonian in the absence of the electric
field, $\vec{F}$.

In this section, we want to show that the displacement operator, 
\[
\hat{T}\left( q\right) =\exp \left[ \frac{i}{\hbar }\left( \left( \vec{q}%
\cdot -i\hbar \nabla _{\vec{r}}\right) \right) \right] , 
\]%
acquires a phase factor in the presence of electric field, and is given by 
\begin{eqnarray}
\tilde{T}\left( \vec{q}\right) &=&\exp \left[ \frac{i}{\hbar }\left( \vec{q}%
\cdot \left( \frac{e}{c}\vec{F}ct\right) +\left( \vec{q}\cdot -i\hbar \nabla
_{\vec{r}}\right) \right) \right] ,  \nonumber \\
&=&\exp \left[ \frac{i}{\hbar }\left( \vec{q}\cdot \left( \frac{e}{c}\vec{F}%
ct\right) +\left( \vec{q}\cdot \hat{P}\right) \right) \right] ,
\label{spaceT}
\end{eqnarray}%
where we indicate by $\tilde{T}$ the translation operator $\hat{T}$ with a
phase factor.

Similarly, for the time displacement operator, we will show that 
\begin{equation}
\hat{T}\left( t\right) =\exp \left[ -\frac{i}{\hbar }\left( \left( t\cdot \
i\hbar \frac{\partial }{\partial t}\right) \right) \right] ,
\label{translationopST}
\end{equation}%
also acquires a phase factor and is given by 
\begin{equation}
\tilde{T}\left( t\right) =\exp \left[ -\frac{i}{\hbar }\left( t\left( e\vec{F%
}\cdot \vec{q}\right) +\left( t\ \mathcal{H}\right) \right) \right] ,
\label{timeT}
\end{equation}%
where $\tilde{T}\left( t\right) $ differs from $\hat{T}\left( t\right) $ by
a phase factor, and $i\hbar \frac{\partial }{\partial t}:\mathcal{=H}$ is
the energy operator of the system. In other words, displacement in space by
lattice vector $q$ acquires phase factors equal to $\exp \left( \frac{i}{%
\hbar }e\vec{F}t\cdot \vec{q}\right) $. Smilarly, displacement in time by $t$
acquires phase factor equal to $\exp \left( -\frac{i}{\hbar }\left( e\vec{F}%
\cdot \vec{q}\right) t\right) $. In the absence of the electric field these
phase factors give unity, e.g., $\tilde{T}\left( t\right) \Longrightarrow _{%
\vec{F}\rightarrow 0}\hat{T}\left( t\right) $, reduces to ordinary
translation operator.

The physics behind these phase factors is dictated by a selfconsistent
translation of \textit{local} functions in space and time under a uniform
electric field, $\vec{F}$. These follow from the self-consistent Heisenberg
equation of motion for quantum operators. For efficient bookeeping and for
ease i taking lattice Weyl transform, it is usually more convenient to
attach these phase factors to displaced local functions themselves, Eq. (\ref%
{peierls}) . This yields a generalization of Peierls phase factor,
well-known for solid-state problems in magnetic fields. The derivation goes
as follows.

\subsubsection{Nonlocality in coordinates}

The nature of the derivatives of exponential displacement operator is
determined, e.g., by the following operation, 
\begin{eqnarray*}
i\hbar \frac{\partial }{\partial t}\tilde{T}\left( q\right) W_{\lambda
}\left( r-0\right) &=&i\hbar \frac{\partial \phi }{\partial t}\tilde{T}%
\left( q\right) W_{\lambda }\left( r-0\right) , \\
\frac{\partial }{\partial t}\tilde{T}\left( q\right) &=&\frac{\partial \phi 
}{\partial t}\tilde{T}\left( q\right) , \\
\frac{d\tilde{T}\left( q\right) }{\tilde{T}\left( q\right) } &=&d\phi ,
\end{eqnarray*}%
where specifically $\hat{T}\left( -q\right) W_{\lambda }\left( r-0\right) $
is a translation of the center coordinate of a localized Wannier function
centered in the origin to another lattice point $q$, yielding $W_{\lambda
}\left( r-q\right) .$ Therefore $\hat{T}\left( -q\right) $ resembles the
physical process of transfering a localized function centered at the origin
to a localized function centered at another lattice point $q$. Here we
define total $\phi $ as undetermined for the moment as 
\[
\phi =-\frac{i}{\hbar }\left( f\left( q,t\right) +\left( \vec{q}\cdot \hat{P}%
\right) \right) , 
\]%
where $f\left( q,t\right) $ is to be determined. Note that the presence of $%
f\left( q,t\right) $ is needed for selfconsistency in the presence of
electric field. Now consider the second term in the exponent, namely, 
\begin{eqnarray*}
\phi _{2} &=&-\frac{i}{\hbar }\vec{q}\cdot \hat{P}=-\frac{i}{\hbar }\vec{q}%
\cdot \left( -i\hbar \frac{\partial }{\partial \vec{r}}\right) , \\
&=&\left( -q\right) \cdot \frac{\partial }{\partial \vec{r}}
\end{eqnarray*}%
leading to Fourier series expansion of the translation, $\hat{T}\left(
-q\right) $ . Then, we obtain 
\begin{equation}
-i\hbar \frac{\partial \phi _{2}}{\partial t}=\left[ H,\phi _{2}\right] =%
\frac{i}{\hbar }\left[ H,\left( \left( -\vec{q}\right) \cdot \hat{P}\right) %
\right] =e\vec{F}\cdot \left( -\vec{q}\right) ,  \label{eFq}
\end{equation}%
since $-\frac{i}{\hbar }f\left( q,t\right) $ commutes with the Hamiltonian.
Therefore, we have 
\begin{equation}
\frac{d\tilde{T}\left( -q\right) }{\tilde{T}\left( -q\right) }=d\phi =\frac{i%
}{\hbar }e\vec{F}\cdot \left( -\vec{q}\right) \ dt.  \label{devlog}
\end{equation}%
We may thus write 
\begin{equation}
-i\hbar \frac{\partial }{\partial t}\tilde{T}\left( \vec{q},t\right) =\left[
H,\tilde{T}\left( \vec{q},t\right) \right] =e\vec{F}\cdot \left( \vec{q}%
\right) \ \tilde{T}\left( \vec{q},t\right) .  \label{eq1}
\end{equation}%
Therefore 
\begin{eqnarray}
\tilde{T}\left( q\right) &=&\exp \frac{i}{\hbar }\left[ \left( e\vec{F}%
\Delta t\right) \cdot \vec{q}\ +\hat{P}\cdot \vec{q}\right] ,  \nonumber \\
&=&\exp \frac{i}{\hbar }\left[ \hat{P}+e\vec{F}t\right] \cdot \vec{q}
\label{spacetransop}
\end{eqnarray}%
This means that a displacement by $\vec{q}$ of localized function acquires a
phase factor given by 
\begin{equation}
Peierls\ phase\ factor=\exp \left[ \frac{i}{\hbar }\left( e\vec{F}t\cdot 
\vec{q}\ \right) \right] .  \label{phsefactorq}
\end{equation}%
We can also deduce from Eq. (\ref{eFq}) the relation for the momentum
operator, 
\begin{eqnarray}
\frac{i}{\hbar }\left[ H,\left[ \left( \vec{q}\right) \cdot -i\hbar \nabla _{%
\vec{r}}\right] \right] &=&e\vec{F}\cdot \left( \vec{q}\right) =\frac{%
\partial \hat{P}}{\partial t}\cdot \left( q\right) ,\Longrightarrow \frac{%
\partial \hat{P}}{\partial t}=e\vec{F},  \nonumber \\
&\Longrightarrow &\hat{P}=\hat{P}_{o}+e\vec{F}t.  \label{canonicalP}
\end{eqnarray}%
which indicates a covariant derivative in spatial coordinates, 
\[
\vec{D}_{\mu }=\vec{\partial}_{\mu }+i\frac{e}{\hbar c}\vec{A} 
\]%
with\bigskip 
\[
\vec{A}_{\mu }=\vec{F}ct 
\]%
as obtained before.

\subsubsection{Simultaneous eigenvalues for $\hat{H}$ and $\tilde{T}\left( 
\vec{q},t\right) $}

One very important conclusion is implied in Eq. (\ref{eq1}), which we
rewrite here for convenience 
\begin{equation}
\left[ \hat{H},\tilde{T}\left( \vec{q},t\right) \right] =e\vec{F}\cdot \vec{q%
}\ \tilde{T}\left( \vec{q},t\right) .  \label{discuss}
\end{equation}%
What the above relation means is that if $\tilde{T}\left( \vec{q},t\right) $
is diagonal then $\left[ H,\tilde{T}\left( \vec{q},t\right) \right] $ is
also diagonal. But if $\tilde{T}\left( \vec{q},t\right) $ is diagonal, then $%
\hat{H}$ is also diagonal with the same eigenvalues. The eigenfunction of $%
\tilde{T}\left( \vec{q},t\right) $ is labeled by the quantum label $\mathcal{%
\vec{K}}=\vec{p}_{o}+e\vec{F}t$. \ This implies that $\hat{H}$ is also
diagonal in $\mathcal{\vec{K}}$. The electric Bloch function labeled by $%
B\left( k_{o}+\frac{e}{\hbar }Ft,..\right) $ is the eigenfunction of $\tilde{%
T}\left( \vec{q},t\right) $ as well as that of the renormalized 
\begin{equation}
H_{renormalized}\left( \hat{K},\hat{Q}\right) \Longleftrightarrow
W_{n}\left( \mathcal{\vec{K}},\mathcal{E}\right) ,  \label{kappaenergy}
\end{equation}%
where the double pointed arrow denotes lattice Weyl correspondence, see Eq. (%
\ref{matrixphase}) below, using the localized electric Wannier function.

We now show the the energy variable $\mathcal{E}$ in Eq. (\ref{kappaenergy})
does incorporate the coordinates through the gauge potential $A_{0}$.

\subsubsection{Nonlocality in time}

Because nonlocal arguments in time acquires phase factor also, the energy
variable of the theory now varies with $e\vec{F}\cdot \vec{q},$ as discussed
next. From Eq. (\ref{translationopST}) for a displacement in time, 
\[
\hat{T}\left( t\right) =\exp \frac{i}{\hbar }\left( -\mathcal{H}t\right) 
\]%
or 
\[
\tilde{T}(t)\equiv \mathcal{F}\exp \left( \frac{-i}{\hbar }\left( i\hbar t%
\frac{\partial }{\partial t}\right) \right) =\mathcal{F}\exp \left( t\frac{%
\partial }{\partial t}\right) 
\]%
where $\mathcal{F}$ incorporates the scalar gauge potential, $A_{0}$. Let 
\begin{eqnarray}
\mathcal{F} &=&\exp \left[ i\alpha ^{0}A_{0}t\right]  \label{timephase} \\
&=&\exp \left[ -\frac{i}{\hbar }\left( f_{t}\left( q,t\right) \right) \right]
,
\end{eqnarray}%
is a phase factor to be determined. The total phase is 
\[
\phi _{t}\left( q\right) =-\frac{i}{\hbar }\left( f_{t}\left( q,t\right) +t\
\left( i\hbar \frac{\partial }{\partial t^{\prime }}\right) \right) . 
\]%
From the equation of motion,%
\begin{eqnarray}
-i\hbar \frac{\partial \tilde{T}(t)}{\partial \vec{q}} &=&-i\hbar \frac{%
\partial \phi _{t}\left( q\right) }{\partial \vec{q}}\tilde{T}(t),  \nonumber
\\
-i\hbar \frac{\partial \phi _{t}\left( q\right) }{\partial \vec{q}} &=&-%
\frac{i}{\hbar }\left[ \mathcal{\vec{K}}\left( t^{\prime }\right) ,t\left(
i\hbar \frac{\partial }{\partial t^{\prime }}\right) \right] =\left[ \vec{p}+%
\frac{e}{c}\vec{A},t\nabla _{t^{\prime }}\right] ,  \nonumber \\
&=&\left[ \vec{p}+e\vec{F}t^{\prime },t\nabla _{t^{\prime }}\right] =-e\vec{F%
}t.  \label{timedispla}
\end{eqnarray}%
Therefore 
\begin{eqnarray*}
-i\hbar \frac{\partial \phi _{t}\left( q\right) }{\partial \vec{q}} &=&-e%
\vec{F}t, \\
\frac{\partial \phi _{t}\left( q\right) }{\partial \vec{q}} &=&-\frac{i}{%
\hbar }e\vec{F}t.
\end{eqnarray*}%
Thus, we obtained, 
\[
\frac{\partial \tilde{T}(t)}{\partial \vec{q}}=\left( -\frac{i}{\hbar }e\vec{%
F}t\right) \tilde{T}(t), 
\]%
and hence,%
\begin{eqnarray*}
d\ln \tilde{T}(t) &=&\left( -\frac{i}{\hbar }e\vec{F}t\right) \cdot d\vec{q},
\\
\ln \tilde{T}(t) &=&\left( -\frac{i}{\hbar }e\vec{F}t\right) \cdot \Delta 
\vec{q}.
\end{eqnarray*}%
Hence a displacement in time carries a phase factor given by $\exp \left( -%
\frac{i}{\hbar }\left( e\vec{F}t\right) \cdot \Delta \vec{q}\right) $ and 
\begin{eqnarray}
\tilde{T}(t) &=&\exp \left\{ -\frac{i}{\hbar }\left[ t\left( e\vec{F}\cdot
\Delta \vec{q}\right) +t\left( i\hbar \frac{\partial }{\partial t^{\prime }}%
\right) \right] \right\}  \nonumber \\
&=&\exp \left\{ -\frac{i}{\hbar }\left[ t\left( e\vec{F}\cdot \vec{q}\right)
+t\left( i\hbar \frac{\partial }{\partial t^{\prime }}\right) \right]
\right\}  \label{timetransop}
\end{eqnarray}%
Once more this gives support to the covariant derivative,%
\[
D_{0}=\partial _{t}+i\alpha ^{0}A_{0} 
\]%
where%
\[
i\alpha ^{0}A_{0}=-i\frac{e\vec{F}\cdot \vec{q}}{\hbar } 
\]%
Therefore, 
\begin{equation}
-i\hbar \frac{\partial }{\partial \vec{q}}\tilde{T}\left( \vec{q},t\right) =%
\left[ \mathcal{\vec{K}}\left( t\right) ,\tilde{T}\left( \vec{q},t\right) %
\right] =\left( -e\vec{F}t\right) \ \tilde{T}\left( \vec{q},t\right)
\label{eq2}
\end{equation}%
Now from second line of Eq. (\ref{timedispla}), we have 
\begin{eqnarray*}
-\frac{i}{\hbar }\left[ \mathcal{\vec{K}}\left( t^{\prime }\right) ,\left(
t\right) i\hbar \frac{\partial }{\partial t^{\prime }}\right] &=&\left[ \vec{%
p}+e\vec{F}t^{\prime },\left( t\right) \frac{\partial }{\partial t^{\prime }}%
\right] =-e\vec{F}t, \\
\left[ \mathcal{\vec{K}}\left( t^{\prime }\right) ,\mathcal{H}\right]
&=&-i\hbar e\vec{F}=-i\hbar \frac{\partial \mathcal{E}}{\partial \vec{q}},
\end{eqnarray*}%
which leads to to the expression for in Eq. (\ref{timephase}) for $f_{t}$ in
the phase factor as 
\[
\frac{\partial \mathcal{E}}{\partial \vec{q}}=e\vec{F}\Longrightarrow f_{t}=e%
\vec{F}\cdot \vec{q}. 
\]%
All these results, Eq. (\ref{spacetransop}) and Eq. (\ref{timetransop}),
lead to the time-dependent wave vector, 
\[
\hbar \vec{k}=\hbar \vec{k}_{o}+\hbar e\vec{F}t 
\]%
and to the position-dependent energy, 
\begin{equation}
\mathcal{E}=E_{o}+e\vec{F}\cdot \vec{q},  \label{energyvar}
\end{equation}%
respectively. Again, for for taking lattice Weyl transform, it is more
convenient to attach these phase factor to the displaced local functions
themselves. This allows us to generalize the Peierls phase factor to space
and time displacements, originally well-known for solid-state problems for
magnetic fields.

As an example, for nonequilibrium translational symmetric and steady-state
condition.%
\begin{equation}
\left\langle q_{1},t_{1}\right\vert \mathcal{O}\left\vert
q_{2},t_{2}\right\rangle \Longrightarrow e^{i\frac{e}{\hbar }\vec{F}t\cdot
\left( \vec{q}_{2}-\vec{q}_{1}\right) }e^{-i\frac{e}{\hbar }\vec{F}\cdot 
\vec{q}\left( t_{2}-t_{1}\right) }O\left( \vec{q}_{2}-\vec{q}%
_{1},t_{2}-t_{1}\right)  \label{peierlsPhase}
\end{equation}%
where%
\[
\vec{q}=\frac{1}{2}\left( \vec{q}_{1}+\vec{q}_{2}\right) 
\]%
\[
t=\frac{1}{2}\left( t_{1}+t_{2}\right) 
\]

\section{Lattice Weyl transforms}

Using the four dimensional notation: $p=\left( \vec{p},E\right) $ and $%
q=\left( q,t\right) $, the lattice Weyl transform (LWT), $A\left( p,q\right) 
$ of any operator $\hat{A}$ is defined by 
\begin{eqnarray}
A_{\lambda \lambda ^{\prime }}\left( p,q\right) &=&\dsum\limits_{v}e^{\left( 
\frac{2i}{\hbar }\right) p.v}\left\langle q-v,\lambda \right\vert \hat{A}%
\left\vert q+v,\lambda ^{\prime }\right\rangle  \nonumber \\
&=&\dsum\limits_{u}e^{\left( \frac{2i}{\hbar }\right) q.u}\left\langle
p+u,\lambda \right\vert \hat{A}\left\vert p-u,\lambda ^{\prime }\right\rangle
\label{LWT}
\end{eqnarray}

Writing Eq. (\ref{LWT}) explicitly, we have%
\begin{equation}
A_{\lambda \lambda ^{\prime }}\left( \vec{p}.\vec{q};E,t\right)
=\sum\limits_{\vec{v};\tau }e^{\left( \frac{2i}{\hbar }\right) \vec{p}\cdot 
\vec{v}}e^{\left( -\frac{i}{\hbar }\right) E\tau }\left\langle \vec{q}-\vec{v%
};t-\frac{\tau }{2},\lambda \right\vert \mathbf{\hat{A}}\left\vert \vec{q}+%
\vec{v};t+\frac{\tau }{2},\lambda ^{\prime }\right\rangle \text{.}
\label{LWT2}
\end{equation}%
Using the form of matrix elements in Eq. (\ref{peierlsPhase}), we have%
\begin{eqnarray}
&&\left\langle \vec{q}-\vec{v};t-\frac{\tau }{2},\lambda \right\vert \mathbf{%
\hat{A}}\left\vert \vec{q}+\vec{v};t+\frac{\tau }{2},\lambda ^{\prime
}\right\rangle  \nonumber \\
&=&e^{i\frac{e}{\hbar }\vec{F}t\cdot \left( \vec{q}_{1}-\vec{q}_{2}\right)
}e^{-i\frac{e}{\hbar }\vec{F}\cdot \vec{q}\left( t_{1}-t_{2}\right) }A\left( 
\vec{q}_{1}-\vec{q}_{2},t_{1}-t_{2}\right)  \nonumber \\
&=&e^{i\frac{e}{\hbar }\vec{F}t\cdot \left( 2\vec{v}\right) }e^{-i\frac{e}{%
\hbar }\vec{F}\cdot \vec{q}\tau }A_{\lambda \lambda ^{\prime }}\left( \vec{q}%
_{1}-\vec{q}_{2},t_{1}-t_{2}\right) \text{.}  \label{matrix}
\end{eqnarray}%
Thus%
\begin{eqnarray}
A_{\lambda \lambda ^{\prime }}\left( \vec{p}.\vec{q};E,t\right)
&=&\sum\limits_{\vec{v};\tau }e^{\left( \frac{i}{\hbar }\right) \vec{p}\cdot
2\vec{v}}e^{i\frac{e}{\hbar }\vec{F}t\cdot \left( 2\vec{v}\right) }e^{\left(
-\frac{i}{\hbar }\right) E\tau }e^{-i\frac{e}{\hbar }\vec{F}\cdot \vec{q}%
\tau }A_{\lambda \lambda ^{\prime }}\left( \vec{q}_{1}-\vec{q}%
_{2},t_{1}-t_{2}\right) ,  \nonumber \\
&=&\sum\limits_{\vec{v};\tau }e^{\left( \frac{2i}{\hbar }\right) \left( \vec{%
p}+e\vec{F}t\right) \cdot \vec{v}}e^{\left( -\frac{i}{\hbar }\right) \left(
E+e\vec{F}\cdot \vec{q}\right) \tau }A_{\lambda \lambda ^{\prime }}\left(
2v,\tau \right) ,  \nonumber \\
&=&A_{\lambda \lambda ^{\prime }}\left[ \left( \vec{p}+e\vec{F}t\right)
,\left( E+e\vec{F}\cdot \vec{q}\right) \right] ,  \nonumber \\
&=&A_{\lambda \lambda ^{\prime }}\left( \mathcal{\vec{K}};\mathcal{E}\right) 
\text{.}  \label{matrixphase}
\end{eqnarray}%
Hence the expected dynamical variables in the phase space including the time
variable occurs in particular combinations of $\mathcal{\vec{K}}$ and $%
\mathcal{E}$. Therefore, besides \ the crystal momentum varying in time as 
\begin{equation}
\mathcal{\vec{K}}=\vec{p}_{o}+e\vec{F}t\text{,}  \label{shiftedK}
\end{equation}%
the energy variable vary with $\vec{q}$ as 
\begin{equation}
\mathcal{E}=E_{o}+e\vec{F}\cdot \vec{q}.  \label{shiftedE}
\end{equation}%
This is the result of covariant derivatives for energy band dynamics in the
presence of uniform electric fields. Thus, differentiation with respect to
coordinate and time variables are now relegated to differentiation with
respect to energy $\mathcal{E}$ and $\mathcal{\vec{K}}$, respectively, 
\begin{eqnarray}
\text{ \ \ }\frac{\partial }{\partial t} &=&\frac{\partial \mathcal{\vec{K}}%
}{\partial t}\cdot \frac{\partial }{\partial \mathcal{\vec{K}}}=e\vec{F}%
\cdot \frac{\partial }{\partial \mathcal{\vec{K}}}\text{,}  \label{dt} \\
\text{\ }\frac{\partial }{\partial \vec{q}} &=&\frac{\partial \mathcal{E}}{%
\partial \vec{q}}\frac{\partial }{\partial \mathcal{E}}=e\vec{F}\frac{%
\partial }{\partial \mathcal{E}}\text{,}  \label{dE} \\
\frac{\partial }{\partial \mathcal{\vec{K}}} &=&\frac{\partial \mathcal{E}}{%
\partial \mathcal{\vec{K}}}\frac{\partial }{\partial \mathcal{E}}=\vec{v}_{g}%
\frac{\partial }{\partial \mathcal{E}}\text{,}  \label{vg}
\end{eqnarray}%
where $v_{g}$ is the group velocity. The LWT of the effective or
renormalized lattice Hamiltonian $\mathcal{H}_{eff}\leftrightarrows H\left( 
\vec{p},\vec{q};E_{o},t\right) $ can therefore be analyzed on $\left( 
\mathcal{\vec{K}},E\right) $-space as%
\begin{equation}
H\left( \vec{p},\vec{q};E_{o},t\right) \Longrightarrow H\left( \mathcal{\vec{%
K}},\mathcal{E}\right) .  \label{HinkappaE}
\end{equation}%
The last line is by virtue of Eqs. (\ref{shiftedK})- (\ref{shiftedE}). Of
course in the absence of the electric field, the dependence in phase space
becomes the familiar $H\left( \vec{p},\mathcal{\omega }\right) $ for
translationally symmetric and steady-state system. But with $\vec{F}\neq 0$
all gauge invariant quantities are functions of $\left( \mathcal{\vec{K}},%
\mathcal{E}\right) $ such as the \textit{electric Bloch function }\cite%
{wannier, bj} or Houston wavefunction \cite{zener}] and \textit{electric
Wannier function}, i.e., the electric-field dependent generalization of
Wannier function. In particular, the Weyl transform of a commutator,%
\begin{equation}
\mathcal{W}\left[ H,G^{<}\right] =\sin \Lambda \ \left\{ H\left( \mathcal{%
\vec{K}},\mathcal{E}\right) G^{<}\left( \mathcal{\vec{K}},\mathcal{E}\right)
\right\} \text{,}  \label{weylcommutr}
\end{equation}%
where $\Lambda $ is the Poisson bracket operator. We can therefore write the
Poisson bracket operator $\Lambda $, as%
\begin{eqnarray}
\Lambda &=&\frac{\hbar }{2}\left[ \frac{\partial ^{\left( a\right) }}{%
\partial t}\frac{\partial ^{\left( b\right) }}{\partial \mathcal{E}}-\frac{%
\partial ^{\left( a\right) }}{\partial \mathcal{E}}\frac{\partial ^{\left(
b\right) }}{\partial t}\right] ,\text{ }  \nonumber \\
&=&\frac{\hbar }{2}\frac{\partial \mathcal{\vec{K}}}{\partial t}\cdot \left[ 
\frac{\partial ^{\left( a\right) }}{\partial \mathcal{\vec{K}}}\frac{%
\partial ^{\left( b\right) }}{\partial \mathcal{E}}-\frac{\partial ^{\left(
a\right) }}{\partial \mathcal{E}}\frac{\partial ^{\left( b\right) }}{%
\partial \mathcal{\vec{K}}}\right] ,  \nonumber \\
&=&\frac{\hbar }{2}e\vec{F}\cdot \left[ \frac{\partial ^{\left( a\right) }}{%
\partial \mathcal{\vec{K}}}\frac{\partial ^{\left( b\right) }}{\partial 
\mathcal{E}}-\frac{\partial ^{\left( a\right) }}{\partial \mathcal{E}}\frac{%
\partial ^{\left( b\right) }}{\partial \mathcal{\vec{K}}}\right] \text{, }
\label{poissonBracket}
\end{eqnarray}%
on $\left( \mathcal{\vec{K}},\mathcal{E}\right) $-phase space.

\section{Application to nonequilibrium quantum transport equation}

The ballistic phase-space quantum transport equation \cite{previous}
generally reads,%
\begin{equation}
\frac{\partial }{\partial t}G^{<}\left( \vec{p},\vec{q};E,t\right) =\frac{2}{%
\hbar }\sin \hat{\Lambda}\left\{ H\left( p,q\right) G^{<}\left( p,q\right)
\right\}  \label{eq105}
\end{equation}%
where in the right side of Eq. (\ref{eq105}) the $4$-dimensional notation of
phase space is employed. Under a uniform electric field, this simplifies to 
\begin{equation}
\frac{\partial }{\partial t}G^{<}\left( \mathcal{\vec{K}},\mathcal{E}\right)
=\sin \Lambda \ \left\{ H\left( \mathcal{\vec{K}},\mathcal{E}\right)
G^{<}\left( \mathcal{\vec{K}},\mathcal{E}\right) \right\}  \nonumber
\end{equation}%
If we expand Eq. (\ref{eq105}) to first-order in the gradient, i.e., $\sin
\Lambda \simeq \Lambda ,$the phase-space transport equation \cite{bj} can be
written in a compact form as%
\begin{equation}
\frac{\partial }{\partial t}G^{<}\left( \mathcal{\vec{K}},\mathcal{E}\right)
=\frac{2}{\hbar }\frac{\hbar }{2}e\vec{F}\cdot \left[ \frac{\partial
^{\left( a\right) }}{\partial \mathcal{\vec{K}}}\frac{\partial ^{\left(
b\right) }}{\partial \mathcal{E}}-\frac{\partial ^{\left( a\right) }}{%
\partial \mathcal{E}}\frac{\partial ^{\left( b\right) }}{\partial \mathcal{%
\vec{K}}}\right] H^{\left( a\right) }\left( \mathcal{\vec{K}},\mathcal{E}%
\right) G^{<\left( b\right) }\left( \mathcal{\vec{K}},\mathcal{E}\right) 
\text{.}  \label{noninteract}
\end{equation}%
With the electric field in the $x$-direction, then we have%
\begin{equation}
G^{<}\left( \mathcal{\vec{K}},\mathcal{E}\right) =e\left\vert \vec{F}%
\right\vert \int dt\ \left[ \frac{\partial ^{\left( a\right) }}{\partial 
\mathcal{\vec{K}}_{x}}\frac{\partial ^{\left( b\right) }}{\partial \mathcal{E%
}}-\frac{\partial ^{\left( a\right) }}{\partial \mathcal{E}}\frac{\partial
^{\left( b\right) }}{\partial \mathcal{\vec{K}}_{x}}\right] H^{\left(
a\right) }\left( \mathcal{\vec{K}},\mathcal{E}\right) G^{<\left( b\right)
}\left( \mathcal{\vec{K}},\mathcal{E}\right) \text{.}  \label{xdirection}
\end{equation}%
The Hall current in the $y$-direction is given by the following equation,%
\[
J_{y}=\frac{a^{2}}{\left( 2\pi \hbar \right) ^{2}}\int \int d\mathcal{\vec{K}%
}_{x}d\mathcal{\vec{K}}_{y}\left( \frac{e}{a^{2}}\frac{\partial \mathcal{E}}{%
\partial \mathcal{\vec{K}}_{y}}\right) \left( -iG^{<}\left( \mathcal{\vec{K}}%
,\mathcal{E}\right) \right) 
\]%
which leads to $J_{y}=\sigma _{yx}\left\vert \vec{F}\right\vert ,$ 
\begin{eqnarray}
\sigma _{yx} &=&e^{2}\frac{1}{\left( 2\pi \hbar \right) ^{2}}\int \int \int d%
\mathcal{\vec{K}}_{x}d\mathcal{\vec{K}}_{y}dt  \nonumber \\
&&\times \ \left[ \frac{\partial ^{\left( a\right) }}{\partial \mathcal{\vec{%
K}}_{x}}\frac{\partial ^{\left( b\right) }}{\partial \mathcal{\vec{K}}_{y}}-%
\frac{\partial ^{\left( a\right) }}{\partial \mathcal{\vec{K}}_{y}}\frac{%
\partial ^{\left( b\right) }}{\partial \mathcal{\vec{K}}_{x}}\right]
H^{\left( a\right) }\left( \mathcal{\vec{K}},\mathcal{E}\right) \left(
-iG^{<\left( b\right) }\left( \mathcal{\vec{K}},\mathcal{E}\right) \right) 
\text{.}  \label{eqKE}
\end{eqnarray}

By reverting to its equivalent matrix element expression and integrating
with respect to time, we obtain%
\begin{equation}
\sigma _{yx}=\frac{e^{2}}{h}\sum\limits_{\alpha }f\left( E_{\alpha }\right) 
\frac{i}{\left( 2\pi \right) }\int \int dk_{x}dk_{y}\left[ 
\begin{array}{c}
\left\langle \alpha ,\frac{\partial }{\partial k_{x}}\mathcal{\vec{K}},%
\mathcal{E}\right\vert \left\vert \alpha ,\frac{\partial }{\partial k_{y}}%
\mathcal{\vec{K}},\mathcal{E}\right\rangle \\ 
-\left\langle \alpha ,\frac{\partial }{\partial k_{y}}\mathcal{\vec{K}},%
\mathcal{E}\right\vert \left\vert \alpha ,\frac{\partial }{\partial k_{x}}%
\mathcal{\vec{K}},\mathcal{E}\right\rangle%
\end{array}%
\right] \text{.}  \label{hall}
\end{equation}%
At low temperature, we can just write Eq. (\ref{hall}) as,%
\begin{eqnarray}
\ \sigma _{yx} &=&\frac{ie^{2}}{2\pi \hbar }\frac{1}{\left( 2\pi \right) }%
\sum\limits_{\alpha }\int \int_{occupiedBZ}dk_{x}dk_{y}\ \left[ \nabla _{%
\vec{k}}\times \left\langle \alpha ,\vec{k}\right\vert \frac{\partial }{%
\partial \vec{k}}\left\vert \alpha ,\vec{k}\right\rangle \right] _{plane}, 
\nonumber \\
&=&\frac{e^{2}}{2\pi \hbar }\frac{i}{\left( 2\pi \right) }%
\sum\limits_{\alpha }\doint dk_{c}\ \left[ \left\langle \alpha ,\vec{k}%
\right\vert \frac{\partial }{\partial k_{c}}\left\vert \alpha ,\vec{k}%
\right\rangle \right] _{contour}\text{.}  \label{lowT}
\end{eqnarray}%
where $\left\langle \alpha ,\vec{k}\right\vert \frac{\partial }{\partial 
\vec{k}}\left\vert \alpha ,\vec{k}\right\rangle $ is the Berry connection in
a band and $\doint dk_{c}\ \left[ \left\langle \alpha ,\vec{k}\right\vert i%
\frac{\partial }{\partial k_{c}}\left\vert \alpha ,\vec{k}\right\rangle %
\right] $ is the Berry phase. Thus, Eq. (\ref{lowT}) yields%
\[
\sigma _{yx}=\sum\limits_{\alpha }\frac{e^{2}}{h}n_{\alpha } 
\]

\section{Concluding remarks}

In this paper, we have shown that the direct route to generalized Peierls
phase factor or Wilson lines for crystalline solid under uniform electric
fields is embodied in the use of covariant derivatives (representing minimal
coupling in $U\left( 1\right) $ gauge theory) in the finite translation
operators. The present calculation thus readily leads to the phase-space
topological invariant in nonequilibrium quantum transport equation which
yields the IQHE of electrical conductivity upon reverting to equivalent
matrix elements expression \cite{previous}.

\begin{acknowledgement}
The author is grateful for the 'Balik Scientist' Visiting Professor grant of
the PCIEERD-DOST, Philippines, at Cebu Normal University, Cebu City,
Philippines.
\end{acknowledgement}


\begin{thebibliography}{99}
\bibitem{previous} Felix A. Buot, \textit{Nonequilibrium superfield and
lattice Weyl transform approach to quantum Hall effect,} arXiv:2001.06993

\bibitem{comments} Felix A. Buot, \textit{Comments on the Weyl-Wigner
calculus for lattice models},http://arxiv.org/abs/2103.10351

\bibitem{jagna} Felix A. Buot, \textit{On the quantization of Hall effect in
electrical conductivity: A nonequilibrium quantum superfield and lattice
Weyl transform transport approach}, AIP Conference Proceedings \textbf{2286}%
, 030007 (2020).

\bibitem{buot8} Felix A. Buot, "Nonequilbrium Quantum Transport Physics in
Nanosystems" (World Scientific, 2009) and references therein.

\bibitem{buot1} F. A. Buot, \textit{Method for Calculating }$TrH^{n}$\textit{%
\ in Solid-State Theory}, Phys. Rev. \textbf{B10}, 3700 (1974).

\bibitem{bj} F. A. Buot and K. L. Jensen, \textit{Lattice Weyl-Wigner
Formulation of Exact Many-Body Quantum Transport Theory and Applications to
Novel Quantum-Based Devices}, Phys. Rev. B\textbf{42}, 9429-9456 (1990).

\bibitem{fego} F.A. Buot, A.R. Elnar, G. Maglasang, and R.E.S. Otadoy, 
\textit{On quantum Hall effect, Kosterlitz-Thouless phase transition, Dirac
magnetic monopole, and Bohr-Sommerfeld quantization}, J. Phys. Commun. 
\textbf{5,} 025007 (2021).

\bibitem{kdp} K. von Klitzing, G. Dorda, and M. Pepper, \textit{A New Method
for High-Accuracy Determination of the Fine--Structure Constant Based on
Quantized Hall Resistance}, Phys. Rev. Lett, \textbf{45}, 494 (1980).

\bibitem{mandel} S. Mandelstam, \textit{Gauge Independent Formulation of
Electrodynamics}, Ann. Phys. (N.Y.) \textbf{19}, 1 (1962).

\bibitem{wu} T.T. Wu and C.N. Yang, \textit{Concept Of Non-integrable Phase
Factors And Global Formulation Of Gauge Fields}, Phys. Rev. D\textbf{12},
3845 (1975).

\bibitem{wannier} G. H. Wannier, \textit{Dynamics of Band Electrons in
Electric and Magnetic Fields}, Rev. Mod. Phys. \textbf{34}, 645 (1962).

\bibitem{zener} F.A. Buot, "Zener Effect", in Encyclopedia of Electrical and
Electronics Engineering, Ed. John Webster, Vol. 23, pp. 669-688 (John Wiley,
NY 1999). Wiley Online Library 2000 John Wiley \& Sons, Inc.
\end{thebibliography}
\end{document}